\documentclass[fleqn,twoside]{article}
\usepackage{espcrc2}

\newcommand{\be}{\begin{equation}}
\newcommand{\ee}{\end{equation}}
\newcommand{\bea}{\begin{eqnarray}}
\newcommand{\eea}{\end{eqnarray}}

\usepackage{graphicx}
\usepackage[figuresright]{rotating}

\title{Dynamical Domain Wall Fermions}

\author{C. Dawson
\address{RIKEN-BNL Research Center,Bldg 510a, Upton, NY 11973-5000}
[RBC Collaboration]
\thanks{We thank RIKEN, Brookhaven National Laboratory and the U.S.\ Department
of Energy for providing the facilities essential for the completion of
this work.}}

\begin{document}

\begin{abstract}
We report on an exploratory study of $N_f=2$ dynamical domain wall fermions
and the DBW2 gauge action at weak coupling. Details of improved simulation
algorithms and preliminary results for the hadron spectrum and renormalised
light and strange quark masses will be presented.
\vspace{1pc}
\end{abstract}

\maketitle

\section{INTRODUCTION}

In the quenched approximation Domain Wall Fermions
\cite{Kaplan:1992bt,Furman:1995ky} (DWF) have been found to be an extremely
successful approach to simulating QCD on the lattice. Crucial to this success
is the fact that, when working at weak couplings ($a^{-1}\approx 2{\rm GeV}$)
and using improved gauge actions such as the DBW2 action, the degree of
explicit chiral symmetry breaking is very small for practically useful sizes
of the fifth dimension ($O(10)$).

Early simulations of dynamical DWF, performed at relatively coarse couplings,
suggested that $L_s \approx O(100)$ would be needed before the degree of
chiral symmetry breaking was small enough to be acceptable.  Here we will
report on the progress of a preliminary study \cite{Izubuchi:2002pt} of
$N_f=2$ dynamical DWF in which we both adopt the DBW2 gauge action, and move
to weaker coupling, in an attempt to find a region of parameter space where
dynamical DWF simulations are practical.

\section{SIMULATION PARAMETERS}

All the results that will be presented were generated using the DWF action
with $L_s=12$ and $M_5=1.8$, and the DBW2 gauge action with $\beta=0.80$ on
$16^3\times32$ lattices. Using the HMC algorithm we have generated three
separate evolutions for bare masses of $m_f=0.02$, $m_f=0.03$ and $m_f=0.04$.
Table~\ref{tab:evo} summarises the total number of trajectories collected so
far, together with the acceptance. Each HMC trajectory is of length 0.5 in HMC
time and is split up into 50 leapfrog integration steps for mass of 0.02 and
0.03, and 40 integration steps for 0.04.
\vspace{-0.5cm}
\begin{table}[!hbt]
\caption{Evolution details}
\label{tab:evo}
\begin{tabular}{ccc}
\hline
$am_f$ & trajectories & acceptance \\
\hline
0.02   & 4716         & 78 \%      \\
0.03   & 4785         & 78 \%      \\
0.04   & 3445         & 68 \%      \\ \hline
\end{tabular}
\end{table}
\vspace{-0.5cm}
\section{ALGORITHMIC DETAILS}

The number of degrees of freedom of DWF grows with $L_s$, but the number of
physical degrees of freedom does not. To cancel off this bulk divergence a
set of Pauli-Villars fields,
\be \Phi^{\dagger} D^{\dagger}(m_f=1)D(m_f=1)
\Phi \, , 
\ee
is added to the DWF Lagrangian for dynamical simulations. Previous work has
used two sets of pseudo-fermion fields to represent the DWF action: one for
the fermion piece of the action and one for Pauli-Villars. The cancellation
between these two terms is therefore only apparent after the average over the
pseudo-fermion fields. Here we have used the fact that
\be \frac{{\rm det} ( D^\dagger(1) D(1)) } { {\rm det} (
D^\dagger(m_f) D(m_f)) } 
\ee
is equal to 
\be {\rm det} ( D(1) [D^\dagger
D(m_f)]^{-1} D^\dagger(1) ) 
\ee
to represent the fermion and Pauli-Villars pieces of the action with a single
pseudo-fermion field. With this approach the cancellation happens step by step
in the leapfrog integration. We find the acceptance of the algorithm is
increased by $10-20\%$ while the inversion costs are reduced by $20-30\%$ when
using this modified force term. 

We have also implemented the chronological inverter technique of
\cite{Brower:1997vx}, leading to a performance improvement of a factor of
$\approx 1.7$. After this improvement calculating a single trajectory
takes approximately $1.6 \times 10^4$, $9 \times 10^3$ and $6 \times 10^3$
dirac matrix applications for $m_f=0.02$, $0.03$ and $0.04$ respectively.

\section{RESULTS}

While the lattices collected represent part of a larger RBC collaboration
project to calculate many hadronic quantities of phenomenological interest,
here we will concentrate on a few mesonic observables to determine the basic
properties of our simulations such as scale and quark mass.  To calculate
these quantities we have used every 50th trajectory, leaving out the first
$\approx 600$ trajectories to allow the evolutions to thermalise. All quoted
errors will be from a jackknife estimate of the statistical error.

To quantify the chiral symmetry breaking from finite $L_s$, we have measured the
residual mass, $m_{\rm res}$ as defined from the breaking term in the Ward-Takahashi
identity \cite{Blum:2000kn}. To extract this we look at 
\be
R(t) = \frac{\sum_{x,y} \langle J_{5q}^a (y,t) \ J_5^a(x,0) \rangle}
{\sum_{x,y} \langle J_{5}^a (y,t) \ J_5^a(x,0) \rangle } \, ,
\ee
which for time greater than some $t_{min}$ should be
time independent and equal to $am_{\rm res}$ \cite{Aoki:2002vt}.
\begin{figure}[!t]
\vspace{-0.4cm}
\begin{center}
\resizebox{7.2cm}{!}{\rotatebox{0}{\includegraphics{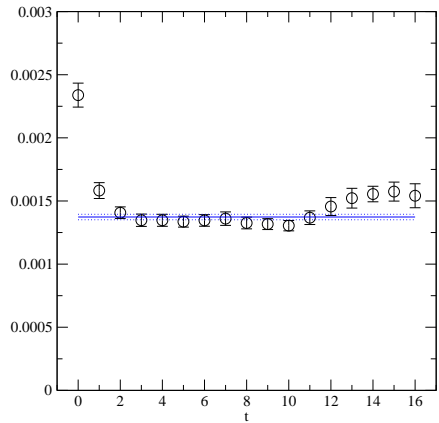}}}
\end{center}
\vspace{-1.5cm}
\caption{residual mass extraction for $m_f=0.02$ evolution}
\vspace{-0.6cm}
\label{MRES_02}
\end{figure}
Figure 1 shows this for the $m_f=0.02$ evolution at the dynamical point. As
can be seen, a plateau is evident for $t \geq 2$, with an error weighted
average between timeslice 6 and the end giving a value of 0.00137(2). This
number is relatively insensitive to the quark mass with linear extrapolation
to $m_f=0$ giving $m_{\rm res} = 0.00136(5)$.

While this value for the residual mass is relatively small compared to the
input quark mass, to properly interpret the value we must know the lattice
spacing and mass renormalisation. For the purpose of this preliminary analysis
we will assign a single lattice spacing for all three evolutions based on a
linear extrapolation in the dynamical quark mass of the $\rho$ meson mass.
\begin{figure}[!t]
\vspace{-0.4cm}
\begin{center}
\resizebox{7.2cm}{!}{\rotatebox{0}{\includegraphics{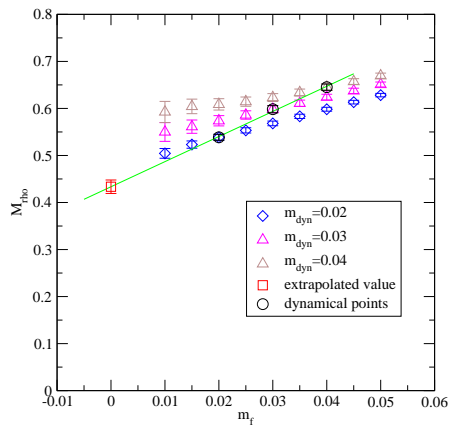}}}
\end{center}
\vspace{-1.5cm}
\caption{$M_\rho$ - dynamical extrapolation}
\vspace{-0.6cm}
\label{SCALE}
\end{figure}
This is shown in Fig~\ref{SCALE}, and leads to an inverse lattice spacing of
$1.806(60) {\rm GeV}$.

\begin{figure}[!t]
\begin{center}
\resizebox{7.2cm}{!}{\rotatebox{0}{\includegraphics{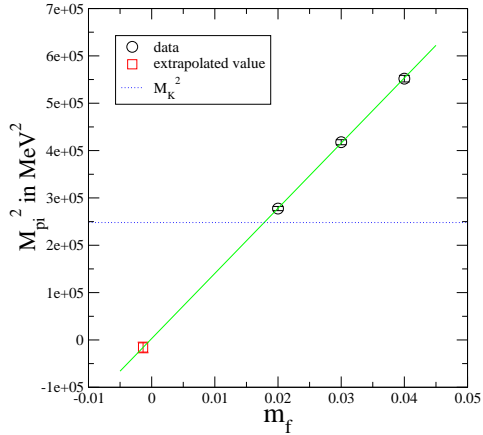}}}
\end{center}
\vspace{-1.5cm}
\caption{$M_\pi^2$ - dynamical extrapolation}
\vspace{-0.4cm}
\label{pi}
\end{figure}
Figure~\ref{pi} shows $M_\pi^2$ for degenerate quark masses versus
dynamical mass. Fitting to the naive, first order chiral perturbation theory,
expectation that
\be M_\pi^2 =
B_\pi \left( m_1 + m_2 \right) \, ,
\label{eq:lin}
\ee
where $B_\pi$ is a constant and $m_1$ and $m_2$ are the quark masses, gives a
$\chi^2$ per degree of freedom of 0.2 and results that are consistent with the
pion mass vanishing at $m_f = -m_{\rm res}$ with $M_\pi^2(m_f=-m_{\rm res}) =
1.6(11) \times 10^4 {\rm MeV}^2$. The experimental value of $M_K^2$ is shown
on the figure as a dotted line. Together with Eq~\ref{eq:lin} this suggests
our lightest quark mass is a little above half the strange quark mass. Going
further, following the same approach as \cite{Dawson:2002nr}, we extract
preliminary values of the renormalised light and strange quark masses of
$3.94(31) {\rm MeV}$ and $103(8) {\rm MeV}$ in the $\overline{MS}$-scheme at
$2 \, {\rm GeV}$.
An alternative way to get a rough idea of the size our input quark mass in
physical terms, which is independent of the way we are setting our scale, is
to calculate the ratio $M_{\pi}/M_{\rho}$. This is 0.541(8), 0.598(8) and
0.637(8) for $m_f=0.02, 0.03$ and $0.04$ respectively.

In quenched simulations using the DBW2 action it was noticed that tunneling
between different topological sectors is suppressed with respect to more
standard gauge actions \cite{Aoki:2002vt}. As such, it is important to 
study how the topological charge varies with trajectory number in our
dynamical simulations.
\begin{figure}[!t]
\begin{center}
\resizebox{7.2cm}{!}{\rotatebox{0}{\includegraphics{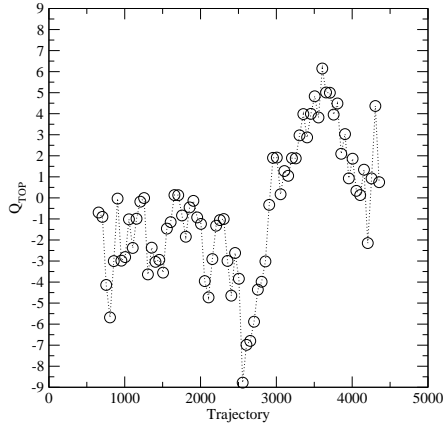}}}
\end{center}
\vspace{-1.5cm}
\caption{Topological charge history for the $m_f=0.02$ evolution.}
\vspace{-0.4cm}
\label{TC_02}
\end{figure}
Fig~\ref{TC_02} shows this for the $m_f=0.02$ evolution, determined using a
classically $O(a^4)$ improved definition of the topological charge calculated
on each lattice after applying 20 steps of APE smearing with a coefficient of
0.45. While it is encouraging that the value of the topological charge is
changing, it is clear that, with a separation of 50 trajectories between
lattices, strong correlations are present.

\section{CONCLUSIONS}

A preliminary study of $N_f=2$ dynamical DWF QCD, using the DBW2 gauge action at an
inverse lattice spacing of $\approx 1.8 {\rm GeV}$, shows that a regime exists
for which the explicit chiral symmetry breaking is small for a computationally
practical extent of the fifth dimension.

\end{document}